\documentclass[twocolumn,a4paper,superscriptaddress,aps,prb]{revtex4-1}

\usepackage{amsmath,wrapfig}			
\usepackage{graphicx}
\usepackage{epstopdf}
\usepackage{mhchem}

\begin{document}    

\title{Revisiting Graphene Oxide Chemistry via Spatially-Resolved Electron Energy Loss Spectroscopy}

\author{Anna Tararan}
\affiliation{Laboratoire de Physique des Solides, Universit\'e Paris-Sud, CNRS UMR
8502, F-91405, Orsay, France}

\author{Alberto Zobelli}
\email{alberto.zobelli@u-psud.fr}
\affiliation{Laboratoire de Physique des Solides, Universit\'e Paris-Sud, CNRS UMR
8502, F-91405, Orsay, France}

\author{Ana M. Benito}
\affiliation{Department of Chemical Processes and Nanotechnology, Instituto de Carboqu\'imica ICB-CSIC, C/Miguel Luesma Cast\'an 4, E-50018 Zaragoza, Spain}

\author{Wolfgang K. Maser}
\affiliation{Department of Chemical Processes and Nanotechnology, Instituto de Carboqu\'imica ICB-CSIC, C/Miguel Luesma Cast\'an 4, E-50018 Zaragoza, Spain}

\author{Odile St\'ephan}
\affiliation{Laboratoire de Physique des Solides, Universit\'e Paris-Sud, CNRS UMR
8502, F-91405, Orsay, France}

\begin{abstract}
The type and distribution of oxygen functional groups in graphene oxide and reduced graphene oxide remain still a subject of great debate. Local analytic techniques are required to access the chemistry of these materials at a nanometric scale. Electron energy loss spectroscopy in a scanning transmission electron microscope can provide  the suitable resolution, but GO and RGO are extremely sensitive to electron irradiation. In this work we employ a dedicated experimental setup to reduce electron illumination below damage limit. GO oxygen maps obtained at a few nanometers scale show separated domains with different oxidation levels. The C/O ratio varies from about 4:1 to 1:1, the latter corresponding to a complete functionalization of the graphene flakes. In RGO the residual oxygen concentrates mostly in regions few tens nanometers wide. Specific energy-loss near-edge structures are observed for different oxidation levels. By combining these findings with first-principles simulations we propose a  model for the highly oxidized domains where graphene is fully functionalized by hydroxyl groups forming a 2D-sp$^3$ carbon network analogous to that of graphane.
\end{abstract}

\maketitle 

\section*{Introduction}

In the last years, graphene oxide (GO) has attracted remarkable interest as a precursor for a large-scale and low-cost production of graphene-based materials and as a fundamental constituent in new functional composite materials for optoelectronics, photovoltaics, and nanobiology devices.\cite{eda_chemically_2010, dreyer_chemistry_2009, loh_graphene_2010, mao_graphene_2012_rev}
GO is obtained by liquid exfoliation of chemically oxidized graphite (graphite oxide);\cite{brodie_atomic_1859,staudenmaier_verfahren_1898,hummers_preparation_1958}
at a later stage, a graphene-like material known as reduced graphene oxide (RGO) can  be derived from GO, by removing the oxygen functionalities.\cite{dreyer_chemistry_2009, pei_reduction_2012_rev}
However, after more than 150 years since its discovery, many questions remain open about the exact chemistry and structure of (R)GO at the atomic scale. \cite{eda_chemically_2010, dreyer_chemistry_2009, loh_graphene_2010, mao_graphene_2012_rev}

Currently, the most acknowledged model for GO corresponds to a nonstoichiometric structure with a random arrangement of epoxide (-O- in bridge position) and hydroxyl (-OH) functionalities at both sides of the carbon basal plane, with small residual nonfunctionalized graphitic areas. \cite{lerf_structure_1998}
Difficulties in defining the exact relative amount and local arrangement of oxygen groups arise from the strong dependence of GO chemistry and structure on the specific graphite precursor and oxidation parameters.\cite{dreyer_chemistry_2009,eda_chemically_2010}
Moreover GO turned out to be unstable  in air and water, resulting in spontaneous reduction and flakes fragmentation.\cite{kim_room-temperature_2012, zhou-105_origin_2013,eigler_formation_2012,dimiev_graphene_2013}
The uncertainties on GO naturally reflect on RGO, whose chemistry and structure also depend on the particular reduction method and parameters.\cite{eda_chemically_2010, dreyer_chemistry_2009, loh_graphene_2010, mao_graphene_2012_rev}
This lack of knowledge represents a strong limit for the understanding of (R)GO physical properties and hence for a controlled use of these promising functional materials.

New insights could be provided only through the use of improved characterization techniques.
Up to now, several spectroscopic techniques have been used to determine the type and amount of oxygen functional groups.
In particular, the oxygen level has been estimated mainly by elemental analysis, XPS and EDX on a variety of samples going from graphite oxide powders to (R)GO films and thin flakes.
The oxygen content has been evaluated in a range between about  15 and 35 atomic percent (at.\%), i.e.  between $\sim$5:1 and $\sim$2:1 in terms of the C/O atomic ratio. In RGO the oxygen content can be lowered to a limit of 0.4 at.\% ($\sim$250:1 C/O atomic ratio), depending on the particular reduction process.\cite{eda_chemically_2010, dreyer_chemistry_2009, loh_graphene_2010, mao_graphene_2012_rev} 
However, these values can only be considered as spatial average quantifications, that cannot account for the irregular chemistry of GO. 
Indeed, strong structural inhomogeneities at the nanometer scale have been revealed in both GO and RGO by high resolution transmission electron microscopy (HRTEM).\cite{wilson_graphene_2009, erickson_determination_2010, gomez-navarro_atomic_2010, pacile_electronic_2011}

Decisive evidence on GO chemistry could only be derived by spectroscopy at the atomic scale. 
Electron energy-loss spectroscopy (EELS) in a scanning transmission electron microscope (STEM) could in principle be suitable for such analysis, providing elemental quantification by core EELS mapping and chemical structure analysis through energy-loss near-edge structures (ELNES), down to the atomic scale.\cite{egerton_electron_2011}
However, the limited use of STEM-EELS  for the study of GO and RGO\cite{mkhoyan_atomic_2009, dangelo_electron_2015} resides in the strong sensitivity of these materials to high energy electron irradiation. 
For instance, atoms mobility and fast amorphization have been clearly demonstrated by time series of HRTEM images, at 80 kV acceleration voltage.\cite{erickson_determination_2010, pacile_electronic_2011} 
Illumination sensitivity is indeed highly expected in the case of GO and RGO, since single or few layer thick specimens and light atoms are intrinsically more affected by radiation damages, such as knock-on of carbon atoms and radiolysis of oxygen functionalities resulting in mass loss.
Irradiation effects in a STEM could in principle be strongly reduced by the use of low accelerating voltages, low electron-dose, and sample cooling. \cite{egerton_control_2013}

In this work, thanks to a dedicated experimental setup combining a liquid nitrogen (LN) cooling system at the sample stage with a low noise LN cooled CCD camera, we investigate few layer GO and RGO by core EELS spectrum imaging in a STEM microscope in low dose acquisition modes. 
We show that the oxygen content in individual (R)GO flakes is strongly heterogeneous at the nanometric scale and that specific near-edge carbon fine structures could be associated with different oxygen contents and bindings, indicating separated chemical phases. 
Finally, on the basis of these experimental evidence and complementary DFT-based numerical simulations, we suggest a structural model for the highly oxidized regions in GO where all carbon atoms are functionalized by -OH groups.  

\section*{Experimental Section}

\paragraph{Synthesis.} Graphene oxide has been synthesized by the modified 
Hummers' method,\cite{hummers_preparation_1958, valles_flexible_2012} and 
successive reduction has been achieved by hydrazine and thermal treatments. Few 
layer flakes with an average lateral size of a few micrometers have been 
obtained by dispersion of the dried material in ethanol and ultrasonication. TEM 
grids were dried in air, first placed in the STEM microscope air lock 
($<2\cdot 
10^{-5}$ mbar), and then moved in the microscope column ($3\cdot 10^{-8}$ mbar) 
and 
cooled down to about 150 K. High angle annular dark field (HAADF) images show a nonhomogeneous  thickness of 
the 
flakes, with thinner regions some hundreds of nanometers wide located at the flakes 
borders. Elemental investigations through core EELS indicate a clear dominance 
of carbon and oxygen  with only negligible residues of  nitrogen (in RGO, 
deriving from the hydrazine reduction agent \cite{stankovich_synthesis_2007, 
valles_flexible_2012}) at few limited regions of the flakes (about 20 nm wide).

\paragraph{Electron Energy Loss Spectroscopy.} EELS spectra were 
acquired in a VG 501 Scanning Transmission Electron Microscope, provided with a 
liquid nitrogen cooling system at the sample stage (150 K). Low temperatures 
have been proven to reduce the mass loss damage by decreasing the atomic 
mobility.\cite{egerton_chemical_1980} Transmitted electrons were collected on a 
liquid nitrogen cooled CCD camera with a low read-out noise of three counts 
r.m.s. and a negligible dark count noise. 
The energy dispersion of the spectra was 0.27 eV, to allow simultaneous 
acquisition of carbon and oxygen K-edges.
The effective energy resolution, resulting from microscope instabilities
and spectrometer aberrations, is estimated to 0.5 eV on the carbon K-edge.
The accelerating voltage has been 
limited to 60 kV, and a low electron dose working mode was set by optimization 
of the acquiring parameters (beam current, illumination area, and time) on (R)GO 
few layer flakes. During spectrum imaging the beam focus (i.e., illumination 
area at the sample stage) has been adjusted to match the pixel size in order to 
avoid oversampling. The exposure of the material to the electron beam before 
spectra acquisition has been prevented by a fast blanking system before the 
sample, following the same procedure as in ref 
\citenum{schooneveld_imaging_2010}. Equally, EELS spectra were acquired prior to 
the corresponding STEM images.

Elemental quantification was performed considering carbon and oxygen edges,
within a 25 eV energy window. Specific cross sections were derived from
the Hartree-Slater model. An error of a few percent may be assumed on the 
quantification results.\cite{bertoni_accuracy_2008}

EELS spectra collected in a low electron dose mode are characterized by little 
intensity of the signals.
Oxygen quantification has been performed on denoised spectra.
Principal component analysis (PCA) represents a common technique for noise 
reduction in core EELS elemental mapping. \cite{de_la_pena-PCA1_mapping_2011} 
PCA analysis has been performed on spectrum images using the HyperSpy 
software.\cite{francisco_de_la_pena_2015_16850} 
After remarking significant changes in carbon fine structure peaks, the number 
of PCA components (less than 10 in the usual routine) was raised to 15.
Fine structure analysis has been performed on the as-acquired spectrum images, 
not treated with PCA, in order to avoid the introduction of any artifacts. 
Nevertheless, in order to appreciate fine structure peaks, a far more intense 
signal with respect to quantification is required.
The signal intensity has been improved by summing spectra arising from 
contiguous areas up to few tens of nanometers wide (pixels in the spectrum 
image) displaying a uniform oxygen content.

\paragraph{DFT Simulations.} Structural optimizations have been 
performed within the framework of density functional theory (DFT) under the local density 
approximation (LDA) as implemented in 
the AIMPRO code.\cite{jones_abinitio_1998, rayson_rapid_2008} Carbon, oxygen 
and hydrogen pseudopotentials are generated using the 
Hartwingsen-Goedecker-Hutter scheme.\cite{hartwigsen_relativistic_1998} Valence 
orbitals are represented by a set of Cartesian gaussians of s-, p-, and d-type 
basis functions centered at the atomic sites. We have used a basis set of 22 
independent functions for carbon (pdpp Gaussian exponents), 40 functions for 
oxygen (dddd), and 12 functions for hydrogen (ppp). The optimization of 
the cell parameter was performed simultaneously with the relaxation of 
the single atomic positions, using a conjugate gradients algorithm.

\section*{Results and Discussion}

\begin{figure*}[tb]\centering
\includegraphics[width=\textwidth]{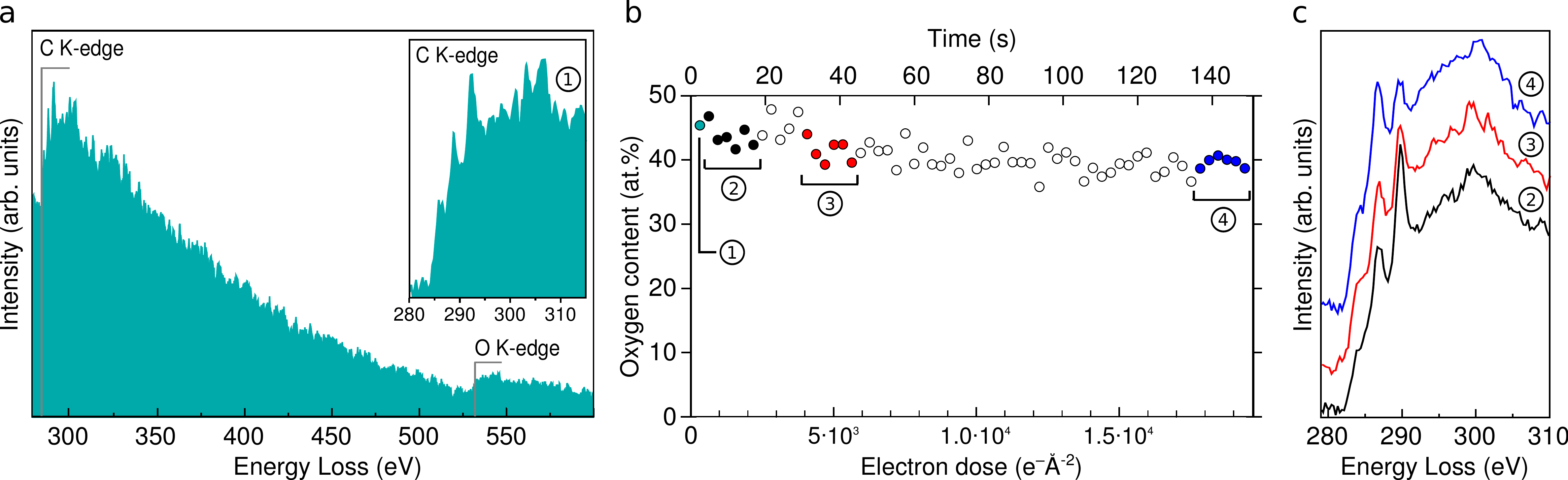}
\caption{Results from a time series of core EELS spectra acquired from
a $44\times44$ nm$^2$ wide thin area of a GO flake. (a) The first 
EELS spectrum, as acquired. (b) Evolution of the oxygen content during 
continuous exposure. The spectrum 
in (a) corresponds to dot (1). (c) Carbon K-edge
spectra corresponding to exposure times indicated by the
integration windows in panel (b).\label{chronospim}} 
\end{figure*}

\paragraph{Optimizing the Dose.}
In order to define optimal conditions for gentle illumination with a 60 keV electron beam, we have monitored the evolution of oxygen content and carbon near-edge fine structures in GO as a function of the irradiating electron dose. In Fig.\ref{chronospim} we present a time series of core EELS spectra acquired under homogeneous illumination over an extended area of a GO flake ($44\times44$ nm$^2$) with a 4 pA electron beam current and an integration time of 2.4 s, corresponding to an electron dose rate as low as 
$1.3\cdot 10^2$ e$^-$\AA$^{-2}$s$^{-1}$. 
The illuminated region has an initial oxygen content of about 45\% (as derived 
from the spectrum in Fig.\ref{chronospim}.a) that progressively reduces as a 
result of the increasing electron dose, as shown in Fig.\ref{chronospim}.b. 
Furthermore, the lowering of the oxygen content is associated with significant 
changes in C K-edge ELNES fine structures ( Fig.\ref{chronospim}.c). Variations 
in the near-edge fine structures are a typical indicator of radiation damages 
affecting chemical bonding and atomic orbital 
hybridization.\cite{egerton_control_2013} 
On this basis, it has been possible to define an upper limit for the electron dose of about $3\cdot 10^3$ e$^{-}$\AA$^{-2}$, below which no substantial changes occur neither in the flakes stoichiometry nor in EELS fine structures. 

The need of low electron dose imposes constraints on STEM spatial resolution.\cite{egerton_control_2013} Indeed, the total acquired spectral intensity is proportional to the electron dose (fixed by irradiation conditions), the elemental specific cross section, the surface projected density of atoms i.e. the number of scattering centers (depending on the sample thickness) and the size of the illuminated area. 
Gentle illumination lowers the intrinsically weak intensity of core EELS signals and high spatial resolution implies that a very low number of scattering centers contribute to the EELS signal. 
In the specific setup of our microscope, the highly efficient CCD camera and the higher electron dose granted by sample cooling\cite{egerton_control_2013} allow for improving the EELS signal intensity. 
The signal/noise (S/N) ratio that is compatible with elemental quantification ultimately defines a $\sim$3 nm wide minimal illumination area for the study of GO.

\paragraph{Elemental Quantification.}

\begin{figure}[tbp]\centering
\includegraphics[width=0.90\columnwidth]{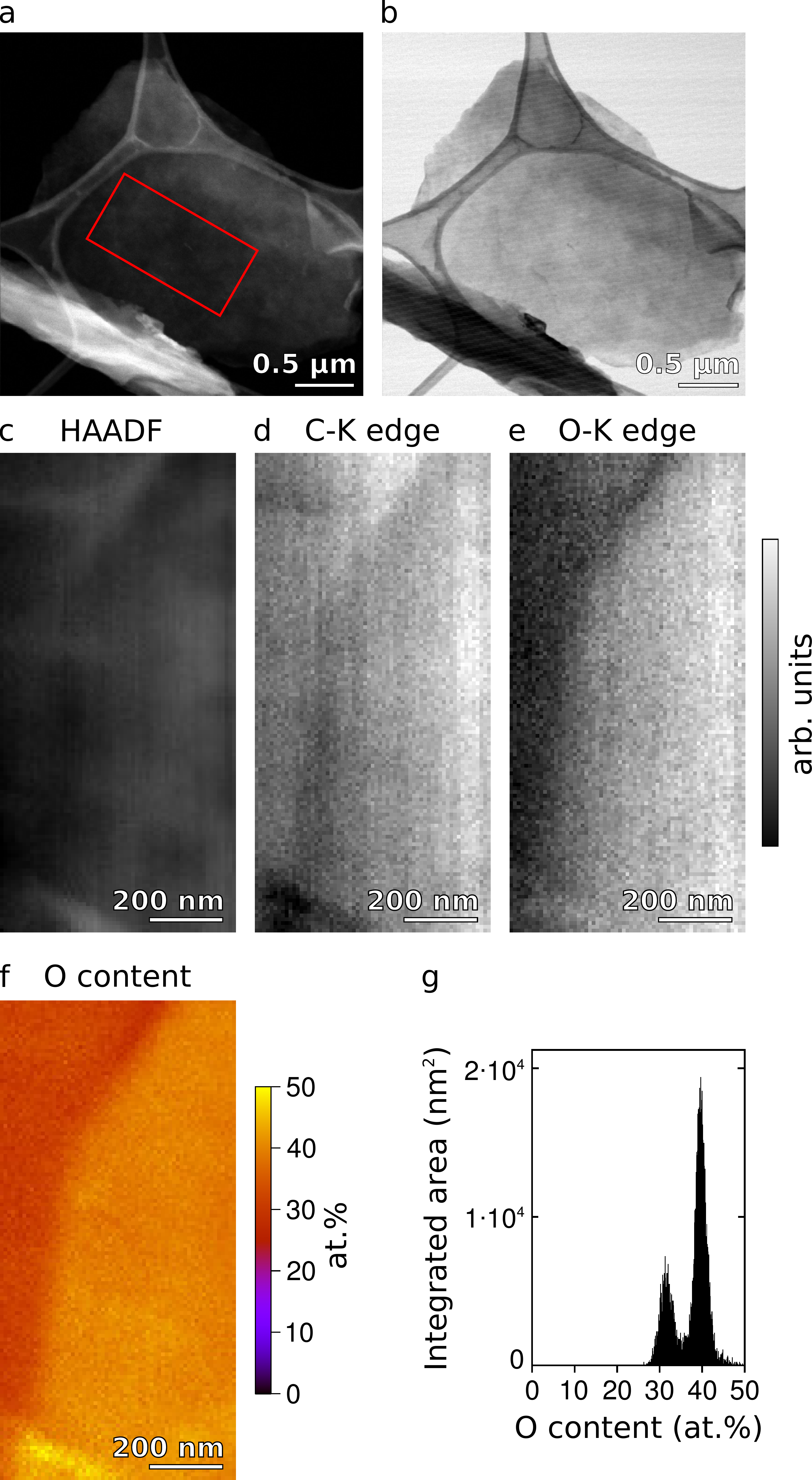}
\caption{Results from a spectrum image on GO. (a) HAADF image and (b) BF image of the whole flake. The spectrum image has been acquired over the region marked by the red rectangle. (c) HAADF image , maps of (d) carbon  and  (e) oxygen  EELS K-edge integrated signals derived from the spectrum image. (f) Map of the relative oxygen content and (g) associated image histogram. \label{carta-GO}} 
\end{figure}

The spatial distribution of oxygen in GO and RGO has been investigated by 
STEM-EELS spectrum imaging. 
This acquisition mode consists in collecting simultaneously the HAADF intensity 
and a complete EELS spectrum at each position of the electron beam during the 
scan, with a $\sim$3 nm minimal beam size and scanning step (no oversampling),
to be compatible with the illumination conditions defined above.
In Fig.\ref{carta-GO} we report an example of EELS oxygen quantification on an 
individual few layer GO flake.
The HAADF and bright field (BF) images of the 
whole flake are presented in Fig.\ref{carta-GO}.a-b, respectively.
A spectrum image has been acquired over the $1.3\times0.6$ $\mu m^2$ wide area 
marked in  Fig.\ref{carta-GO}.a and the resulting HAADF image is shown in  Fig.\ref{carta-GO}.c.
Acquisition parameters were 10 nm spatial resolution (probe size and scanning 
step), a 15 pA current and 0.15 s integration time, that correspond to an 
$\sim$$1.3\cdot 10^3$ e$^{-}$\AA$^{-2}$ total electron dose per spectrum.
The HAADF intensity depends on both the amount of matter (thickness and areal 
density) and the chemical species (atomic number).
This information appears separated in the integrated K-shell ionization edges 
maps derived from the spectrum image ( Fig.\ref{carta-GO}.d-e), that show the 
spatial variation of the surface projected amount of carbon (hence the number of 
carbon layers) and oxygen, respectively.
In the quantification map of oxygen relative to carbon ( Fig.\ref{carta-GO}.f)  
two well separated domains with oxidation levels centered at $\sim$32  at.\% and  
$\sim$40 at.\% are visible, as confirmed by the associated image histogram in  Fig.\ref{carta-GO}.g. 
A third less extended domain (bottom left corner in the map) corresponds to $\sim$45 
at.\% with maximum local content of almost 50 at.\%. 
When interpreting these results it must be considered that for each pixel the EELS 
signal is integrated over the specimen thickness. 
Hence, a nonhomogeneity in the relative quantification map can result from  a 
spatial variation of the local oxidation level in regions with a constant number 
of layers or from the local superposition of additional layers with a different 
oxidation level. 
In  Fig.\ref{carta-GO}.f an increase of the oxygen relative content is visible 
from the left to the right side of the map forming two well separated regions.
 The oxygen integrated K-edge map ( Fig.\ref{carta-GO}.e) shows the same trend: 
the right side area does contain a higher amount of oxygen. 
On the contrary, the intensity of the carbon integrated K-edge ( Fig.\ref{carta-GO}.d) is almost uniform, indicating that the flake has a constant 
carbon thickness along the region displaying the oxygen gradient. 
Therefore these results show a variation of the oxidation level within an 
individual flake, forming spatially well separated oxidation phases which are 
not correlated with the flake thickness.

Overall the oxygen content in GO has been observed to vary between $\sim$10 
at.\% and $\sim$50 at.\% (C/O ratio of 9:1 and 1:1, respectively).
The higher oxidation level (30-50 at.\%) seems to characterize about one-quarter of the 
analyzed material,
while in extremely rare cases the oxygen content falls below the detection limit (unoxidized flakes), 
confirming the 
heterogeneity of GO.
Therefore despite the fact that the local oxygen concentration can be extremely 
high, the average oxygen content in individual flakes is $\sim$25 at.\%, which 
is in good agreement with spatially averaged values reported in the literature.\cite{eda_chemically_2010, dreyer_chemistry_2009, loh_graphene_2010, 
mao_graphene_2012_rev} 
Regions up to a few hundred nanometers wide are characterized by an almost 
uniform oxidation level (at a 3-10 nm scale), while the transition between two 
oxidation phases occurs in the space of few tens of nanometers. This observation 
is compatible with a recently proposed oxidation mechanism where the reaction 
progresses within graphite flakes in a front-like diffusive-controlled 
pathway.\cite{dimiev_mechanism_2014}

\begin{figure}[tbp]\centering
\includegraphics[width=0.90\columnwidth]{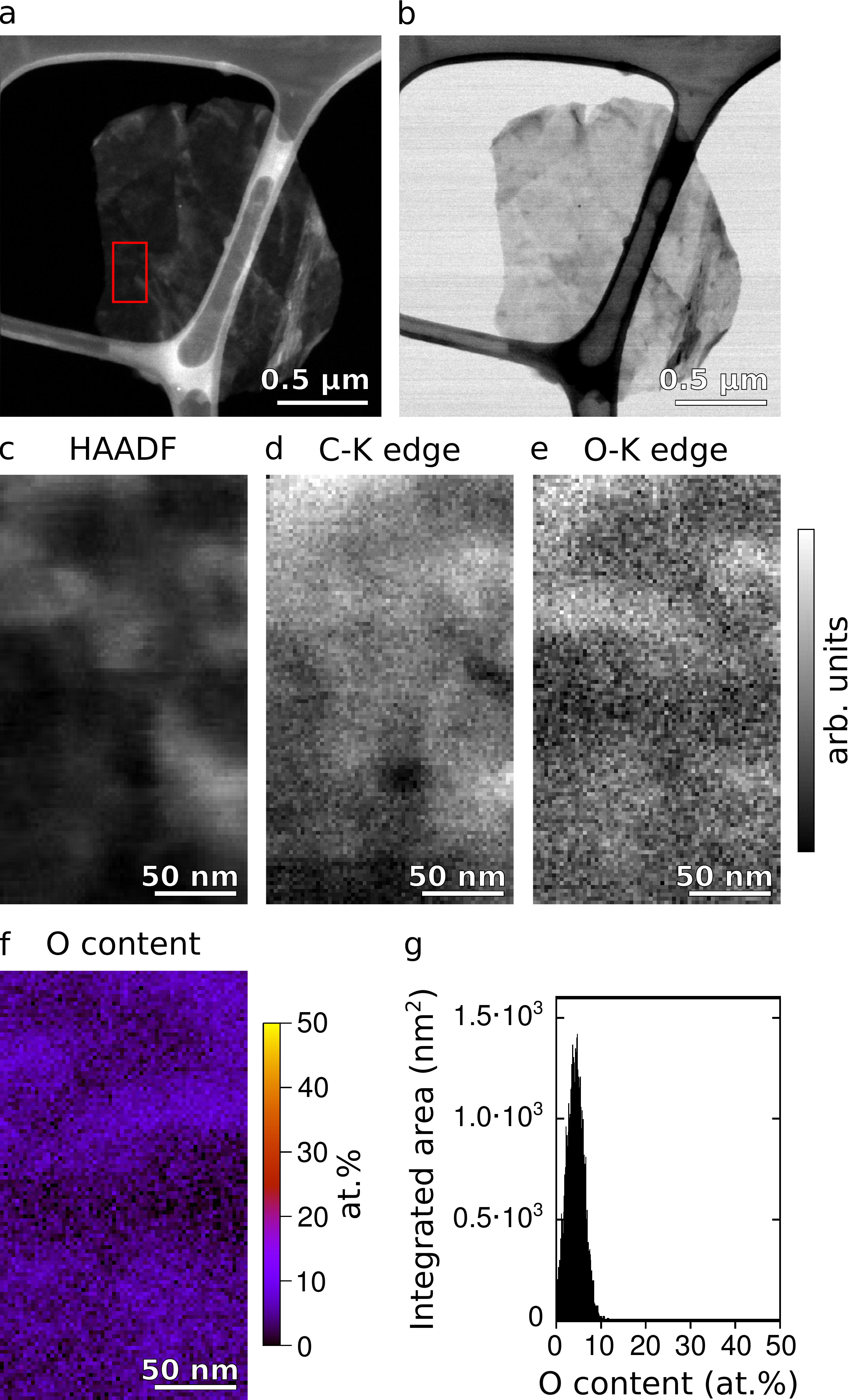}
\caption{Spectrum imaging on a RGO flake. (a) HAADF image and  (b) BF 
image of the whole flake. The region investigated in the spectrum image is
marked by a red rectangle. (c) HAADF image, maps of (d) carbon  and  (e) 
oxygen  EELS K-edge integrated signals derived from the spectrum image. (f) Map 
of the relative oxygen content with (g) associated image histogram. 
\label{carta-RGO}} 
\end{figure}

STEM images and spectroscopic analysis from RGO are shown in  Fig.\ref{carta-RGO}.
HAADF and BF images of the whole flake are shown in  Fig.\ref{carta-RGO}.a-b.
A spectrum image has been collected over the region marked in  Fig.\ref{carta-RGO}.a, over an area of about $330\times190$ nm$^2$. 
Acquisition parameters were as follows: 3 nm spatial resolution, 11 pA current, and 0.04 s 
integration time, for a total electron dose per spectrum of $\sim$$3.0\cdot 
10^3$ e$^{-}$\AA$^{-2}$.
 Fig.\ref{carta-RGO}.c-g shows the associated HAADF image, the integrated 
K-shell ionization edges maps of carbon and oxygen, the quantification map of 
oxygen relative to carbon, and its associated image histogram.
As in GO, the carbon and oxygen K-edge integrated intensity maps appear 
mainly noncorrelated. Weak
correlations appear occasionally, not systematically, in some limited areas.
 Namely, in the central region of the spectrum image the carbon 
K-edge intensity is almost constant, while the oxygen K-edge signal varies and 
may be responsible for the contrast variation in the HAADF image. 
The distribution of oxygen forms patches few tens of nanometers wide ( Fig.\ref{carta-RGO}.e-f) that correspond to an unimodal distribution centered at 
about 5 at.\% ( Fig.\ref{carta-RGO}.g), with a local upper limit of 12 at.\% 
(local residual oxygen up to 20 at.\% has been observed in other flakes, over 
limited regions).
These values are situated in the range of spatially averaged oxygen 
quantifications reported in the literature.\cite{eda_chemically_2010, 
dreyer_chemistry_2009, loh_graphene_2010, mao_graphene_2012_rev}

\paragraph{Core EELS Fine Structures.}

\begin{figure}[tbp]\centering
\includegraphics[width=0.9\columnwidth]{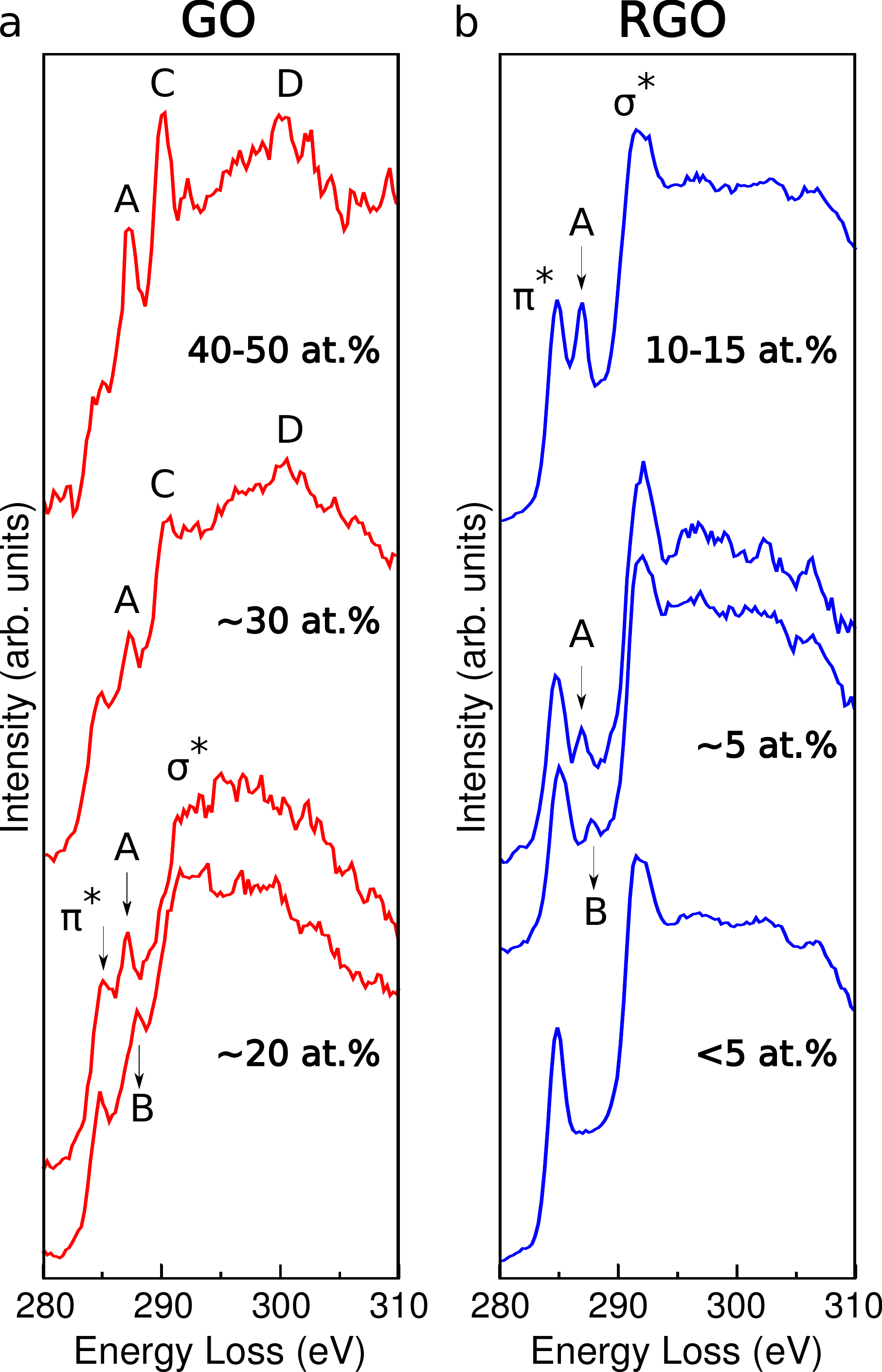}
\caption{Carbon K-edge EELS spectra of GO (a) and RGO (b) from homogeneous 
regions of the flakes, corresponding to an oxygen content comprised in the indicated range.\label{FineStructure-C}} 
\end{figure}

Besides oxygen quantification, coordination states and chemical bonds for 
different oxidation phases can be extracted from spectrum images through 
Energy-Loss Near-Edge Structures analysis. 
The spatial resolution of the spectrum images has been optimized for elemental 
analysis; the higher S/N ratio required for fine structure investigation can be 
obtained by integrating the spectrum images over regions few tens of nanometers 
wide, having a uniform oxygen content. 
In  Fig.\ref{FineStructure-C} we present fine structures at the carbon K-edge 
that are representative of particular ranges of oxygen content.
 Fig.\ref{FineStructure-C}.a shows the fine structure signatures observed in GO 
areas characterized by 10 to 50 oxygen at.\%. 
In both spectra corresponding to low oxygen concentration ($\sim$20 at.\%), the $\pi^*$ and $\sigma^ *$ signatures associated with sp$^2$ carbon are 
visible respectively at 285.0 eV and 292.7 eV.\cite{el-barbary_electron_2006}
Two additional peaks at 287.2 (A) and 288.2 eV 
(B) could be isolated over two distinct areas of GO (in this 
particular case showing 20 and 25 oxygen at.\%, respectively). 
These features generally show a feeble intensity and can appear 
simultaneously in the same spectrum, resulting in mixed features that can hardly 
be identified.
In GO regions containing $\sim$30 oxygen at.\%, the 
intensity of the $\pi^*$ peak lowers, while the 287.2 eV peak (A) 
remains intense (32 oxygen at.\% for the here reported spectrum). 
An additional feature rises at 290.1 eV (C) followed by a broad 
triangular-shaped signal at about 300 eV (D). 
For higher oxidized GO regions (40-50 at.\%, the reported spectrum corresponding to 40\%), the $\pi^*$ peak further lowers, while the peaks at 287.2 eV (A)
and 290.1 eV (C) get stronger and very well-defined.
The 290.1 eV peak appears particularly sensitive to irradiation damages, as shown by  Fig.\ref{chronospim}.c.

After reduction ( Fig.\ref{FineStructure-C}.b) the sharpness of the $\pi^*$ 
and $\sigma^*$ peaks, with a trace of $\sigma^*$ excitonic feature at 291.6 eV \cite{ma_core_1993,bruhwiler_pi_1995}, reveals a very 
good recovery of the graphitic network. 
An additional well-defined peak at 287.2 eV (A) is observed (the two spectra shown correspond to 7 oxygen at.\% and 12 at.\%). 
Rarely and in some limited regions, other weak fine structure peaks have been 
distinguished, i.e. peak B (5 oxygen at.\%). 
These A and B spectral features are below the detection limit for lower oxygen 
concentrations.

These results show that specific near-edge fine structures and thus coordination 
states of the carbon atoms correspond  to different oxidation rates.
The energies of EELS peaks in  Fig.\ref{FineStructure-C} are compatible with 
previous XANES results which however could not separate the specific near-edge 
fine structures shown here due to  spatial averaging over wide regions with 
different oxygen levels. Absorption peaks indicated as A-C had been attributed 
in the literature to {C-O} $\pi$ symmetry antibonds: lower energy peaks to 
single bonds and higher energy peaks  to double bonds. More precisely, hydroxyl 
groups had been generally associated with peak 
A, epoxide groups to 
peak A or B, and carbonyl and carboxyl groups 
to peak B or C.\cite{jeong_x-ray_2008, ganguly_probing_2011,zhan_electronic_2011, 
zhou_nano-scale_2011, lee_soft_2012,lee_quantum_2013, chuang_effect_2014}
In the 
$\sigma^*$ region, carbonyl groups have been related to peak 
D.\cite{ganguly_probing_2011,pacile_electronic_2011} 
In a recent STEM-EELS investigation a shallow peak A has been observed at the 
carbon K-edge and assigned to epoxide, while no peak C has been 
detected.\cite{dangelo_electron_2015}
These assignments, still largely debated, have been obtained by comparison with 
reference spectroscopic signatures of aromatic 
molecules\cite{francis_inner-shell_1992,christl_c-1s_2007} or spectra 
simulations on hypothetical GO atomic structures.\cite{hunt_re-evaluation_2014, 
dangelo_electron_2015}  
However, as shown a number of times by numerical simulations, 
\cite{boukhvalov_modeling_2008, lahaye_density_2009, 
ghaderi-114_first-principle_2010, wang-103_stability_2010, lu_structure_2011} 
the spatial configuration of functional groups within (R)GO deeply affects their 
orbital associated energies, and these might strongly differ from those of the 
same groups within molecules.
Moreover, the sensitivity of the material suggests the need of caution while 
comparing spectroscopic results. In particular, the illumination conditions can 
induce changes in the relative intensity of fine structure peaks by 
transformation of the functional groups (possibly hydroxyls into epoxides) or 
selective sputtering of the oxygen groups.

\paragraph{Structural Model.}

The structure of GO can be revisited on the basis of the here reported spatially 
resolved spectroscopic results.
Several complex atomic models for GO have been proposed in the last 20 years.  \cite{eda_chemically_2010, dreyer_chemistry_2009, loh_graphene_2010, 
mao_graphene_2012_rev}
Oxygen adsorption is usually described in terms of the basic functional groups 
of epoxide, hydroxyl, carbonyl (-CO) and carboxyl 
(-COOH).
The last two groups involve multiple bonding with a single carbon atom, which 
can occur at edges or at defects within the graphene lattice.
On the contrary, a complete and in principle reversible functionalization of a 
perfect graphene network can be achieved with epoxides and hydroxyls, resulting 
in a C/O ratio of 2:1 (33 oxygen at.\%) for solely epoxides functionalization, 
1:1 (50 oxygen at.\%) for solely hydroxyls functionalization, and intermediate 
values for mixed structures.
Up to now, the Lerf-Klinowski model\cite{lerf_structure_1998}  has been the most 
referred to. 
In this representation, different oxygen groups are randomly distributed on both 
sides of the carbon sheets, with coexisting graphitic and oxidized regions, thus 
forming a partially functionalized structure.
Hydroxyls and epoxides are supposed to be the dominant groups on the basal plane, 
while carboxyls are expected to fill carbon dangling bonds at the sheet edges 
and vacancy sites.\cite{lerf_structure_1998}
Carbonyl groups as ketones and/or quinones are also generally expected at the 
plane edges. 
A high proximity among hydroxyls and epoxides 
and the presence of five- and six-membered lactols and ester carbonyls have been 
proposed on the basis of NMR studies.\cite{casabianca_nmr-based_2010, 
gao_new_2009}
With respect to our experimental observation, the Lerf-Klinowski model is 
consistent with GO regions with a 20 oxygen at.\% content, but cannot account for 
the much higher oxygen concentration locally observed by EELS ($\sim$45 at.\%).

The presence of water molecules in the atomic structure of GO is still under 
debate.
Water molecules are expected to form hydrogen bonds with oxygen functional 
groups on GO.\cite{lerf_structure_1998}
We expect reasonably that any possible intercalated water would have been removed from the here
observed flakes,
during the preparation process.
Indeed, differently from the common protocol where GO is directly deposited on a substrate as a water dispersion, 
in this work we employed dried GO which was successively redispersed 
in ethanol and ultrasonicated. 
Moreover observation conditions involve high vacuum levels ($3\cdot 10^{-8}$ mbar), resulting in
evaporation of residual water molecules.
On the basis of these considerations we expect a negligible amount of 
water.
This hypothesis is corroborated by our observed EELS fine structures.
Indeed, the strong intensity and sharpness of the fine structure peaks at the 
carbon K-edge correlated with a high oxygen 
content ( Fig.\ref{FineStructure-C}.a) suggest that a significant amount of 
carbon atoms is involved in
a specific kind of carbon-oxygen bond.
Thus, the $\sim$45 oxygen at.\% in GO cannot arise from intercalated water 
and should be ascribed to oxygen directly bound to carbon.

Recently it has also been proposed that the GO surface could be decorated by 
strongly bound highly oxidized debris coming from the oxidation process. 
\cite{rourke_real_2011}
In the present work, an important contribution of such debris to the calculated 
oxygen content is not compatible with our spectroscopic investigations: their 
presence would result in an increase of intensity in the K-edge maps of both 
oxygen and carbon (as an additional GO layer) and hence in a correlation 
between the two maps, which is not observed in  Fig.\ref{carta-GO}.d-e.
Moreover, the highly oxidized residues could contain at most $\sim$50 oxygen 
at.\%, considering a fully functionalized structure
 (hydroxyls on the basal plane and carbonyls-carboxyls at the edges, as 
discussed above). 
 In  Fig.\ref{carta-GO}.g the local  oxygen content reaches almost 50 at.\% and 
represents an integrated measurement over the flake thickness: even assuming the 
presence of debris, the underlying GO layer(s) has to be necessarily almost 50 
at.\% oxidized.
This experimental evidence reinforces the criticism recently raised against the 
debris GO model. \cite{dimiev_contesting_2015}

\begin{figure}[tb]
\centering
\includegraphics[width=0.9\columnwidth]{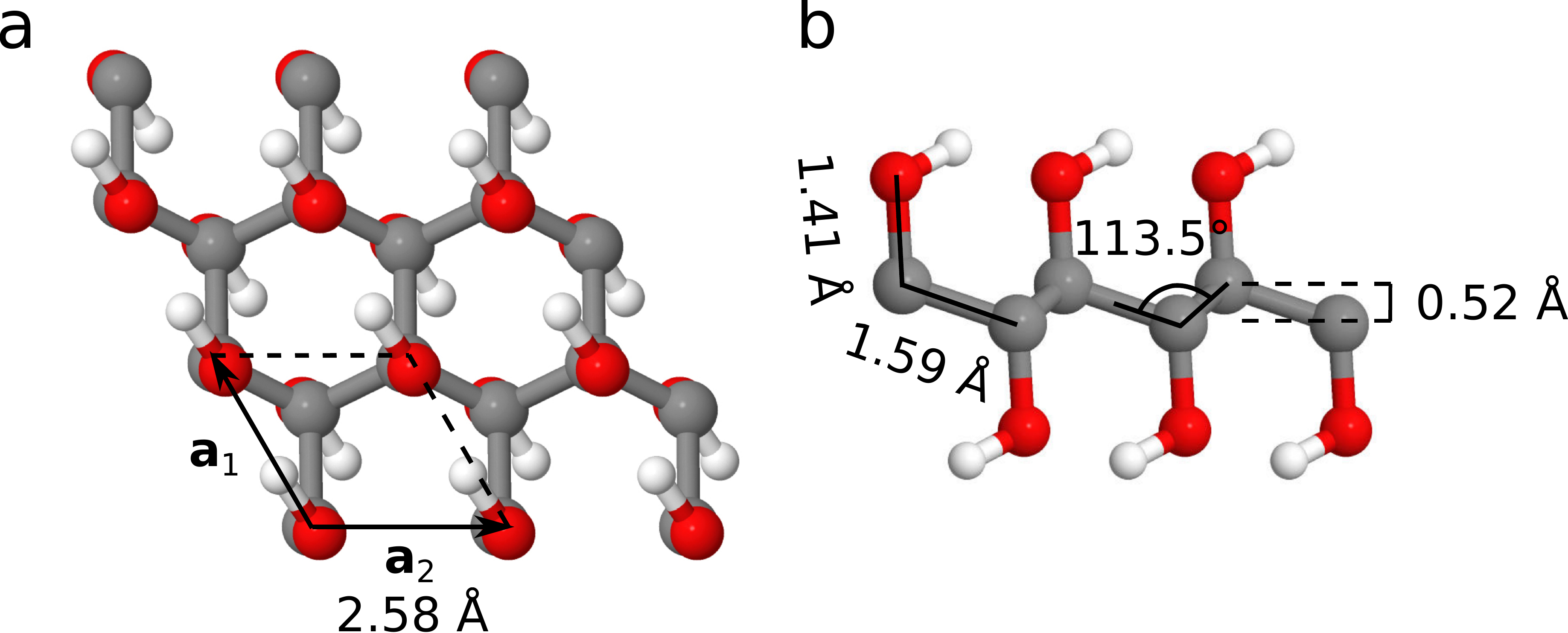}
\caption{Limit structure allowing a C/O ratio of 1:1. Top (a) and side (b) view. 
The indicated structural parameters have 
been obtained after full DFT-LDA relaxation.}
\label{model} 
\end{figure}

First principle simulations have shown that in GO there is a strong driving 
force toward a phase separation in fully functionalized regions (epoxides 
and/or hydroxyls) and pristine graphene regions.\cite{yan_structural_2009, 
wang-103_stability_2010,ghaderi-114_first-principle_2010, zhou-105_origin_2013} 
The limit of 50 at.\% local 
oxygen concentration observed in this work is not consistent with a full 
functionalization dominated by epoxides, since the oxygen content would tend to 
33 at.\%.
A large presence of carboxyl and carbonyl groups should also be excluded, 
because such a highly defective structure  could hardly be reduced to a highly 
graphitic arrangement compatible with the carbon K-edge fine structures observed 
in RGO flakes. 
GO containing almost 50 at.\% of oxygen and converting into a highly graphitic 
structure when reduced can be explained only by an almost hydroxyl saturated 
graphene lattice. This limit structure is analogous to graphane\cite{elias_control_2009} (i.e., hydrogen 
saturated graphene), and hydroxyl groups should 
alternate at the two sides of the carbon plane in order to minimize 
strain.\cite{boukhvalov_modeling_2008, yan_oxidation_2010} This limit 
configuration is presented in  Fig.\ref{model} after optimization by DFT-LDA. 
The graphene network is shown to be strongly distorted, with carbon atoms puckered out 
of plane by about  0.26 \AA\, leading to a change of their hybridization state 
from pure sp$^2$ to partial sp$^3$. In contrast with graphane, where the sp$^3$ 
carbon hybridization reduces the in plane cell parameter ($\sim$2.42 
\AA),\cite{elias_control_2009} we obtain an increased cell parameter of  2.58 
\AA, (about 5\% higher than DFT-LDA optimized graphene) and a C-C bond length higher than diamond 
(1.59 \AA). 
Oxygen steric hindrance and hydrogen bonding between neighboring groups forming 
hydroxyl chains are responsible for this in plane expansion as for graphene 
edges functionalized by hydroxyl groups where the accumulated strain is released 
via the formation of static out-of-plane ripples.\cite{wagner_ripple_2011} 
It might be challenging to directly observe this lattice expansion through 
diffraction because uniform and continuous highly functionalized domains are of 
little extension. 
Furthermore, the accurate determination of the unit cell 
parameter by electron diffraction is a nontrivial technique itself, for in 2D 
materials the reciprocal space of monolayers consists of rods and thus the total 
intensity of the diffraction spots changes little as a function of the tilt angle, 
while the spots broadening is important. \cite{meyer_structure_2007}

EELS fine structures for highly oxidized GO (30-50 oxygen at.\%,  Fig.\ref{FineStructure-C}.a) can
thus be interpreted in the limit of this hydroxyl 
saturated graphene model. 
The extremely low intensity of the  $\pi^*$ peak can be ascribed to the 
loss of sp$^2$ hybridization at carbon atoms.  
The 287.2 eV peak (A), which is observed also for lower oxygen rates, can be 
assigned to {C-OH} antibonds, in  agreement with previous XANES 
works.\cite{lee_soft_2012,lee_quantum_2013,chuang_effect_2014} 
The peak at 290.1 eV (C) cannot be associated with {C-O} antibonds of 
other oxygen functional groups, because they are expected to be present only in 
a small amount and this is not compatible with the strong intensity of the peak.
Nevertheless, the interpretation of this feature remains unclear.

Fine structures for lower oxidation rates (10-30 oxygen at.\%,  Fig.\ref{FineStructure-C}.a) are much less defined. 
This can be due to a higher variety of functional groups and the high number of 
possible configurations that they can assume at the graphene network. 
The presence of the  $\pi^*$ peak and of a weak  sp$^2$ $\sigma^*$ peak infers a limited presence of nonfunctionalized carbons. 
Aside from peak A ascribed to hydroxyl groups, peak B can be attributed to 
epoxide groups. 
After reduction the very intense and sharp peak A observed at higher residual 
oxygen regions ( Fig.\ref{FineStructure-C}.c, $\sim$15 oxygen at.\%) would then 
correspond to the presence of solely hydroxyl groups. 
This interpretation is compatible with the identification of several simple 
reaction routes for epoxide reduction by hydrazine, while hydroxyl removal 
mechanisms are still 
debated.\cite{stankovich_synthesis_2007,gao_hydrazine_2010} 
Considering the same assignments as in GO, the weak peak B in RGO indicates a 
small residual amount of epoxides.
Again, the difference in the intensity of features A and B can derive from their 
relative amount (hydroxyl probably remain the dominant species also after 
reduction) and the different number of configurations that these functional 
groups can form at the carbon network.

\section*{Conclusion}

In conclusion we have investigated the oxygen distribution in GO and RGO by core 
EELS spectroscopy in a STEM microscope. A spatial resolution of 3 nm for spectrum 
images has been achieved by reducing radiation damage through optimized 
acquisition conditions and an experimental setup combining a low noise CCD 
detection camera with a liquid nitrogen cooled sample stage. 
Previous spectroscopic quantifications have provided only overall oxygen 
contents,  and GO and RGO have been often described as chemically homogeneous 
materials at a nanometric scale. 
We have shown that within individual flakes the oxygen content strongly varies, 
forming homogeneous domains with a lateral side of few tens of nanometers.
In GO different oxidation levels have been identified, with a maximum local 
concentration of almost 50 at.\% of oxygen, corresponding to a C/O ratio of 1:1. 
In RGO the residual oxygen mainly concentrates in domains with about 5 oxygen 
at.\%.
Specific fine structures at the carbon K-edge can be clearly associated with 
different oxidation phases. 
Highly oxidized GO displays a shallow graphitic signature with extremely intense 
and sharp additional peaks, not observed before.
In RGO the graphitic character is extensively recovered with weakened additional 
features.
The assignment of the spectroscopic peaks is still debated in the literature, and the 
previously proposed mixed epoxides-hydroxyls atomic model for GO cannot account 
for the highest oxygen concentrations observed in this work. 
We propose a limit structural model for these highly oxidized regions, in which 
all carbon atoms are functionalized with {-OH} groups, leading to an sp$^3$ 
carbon 2D material analogous to graphane. A strong signature for {-OH} antibonds
 can then be identified in GO carbon near-edge  fine 
structures. 
In RGO, a sharp {-OH} antibond peak, localized at highest oxidized regions, 
infers that this group is the dominant surviving species. 

The presence of domains fully covered with hydroxyl groups has critical 
implications for the engineering of the electronic structure of individual 
flakes. Indeed, it strongly affects the hydrophilicity of the flakes and the 
interaction with molecules from the gas and vapor 
phase.\cite{rezania_hydration_2014} Furthermore, the presence of a dense 
hydroxyl layer on (R)GO sheets has important  consequences for the anchoring of 
metal and ceramic nanoparticles or biomolecules and the assembly of individual 
flakes into corresponding macroscopic forms.\cite{nunez_integration_2014}

\section*{Acknowledgements}
A.T., A.Z. and O.S. acknowledge support from the Agence Nationale de la Recherche 
(ANR), program of future investment TEMPOS-CHROMATEM (No. ANR-10-EQPX-50). The 
work has also received funding from the European Union in Seventh Framework 
Programme (No. FP7/2007 -2013) under Grant Agreement No. n312483 (ESTEEM2). A.M.B. 
and W.K.M. are grateful for financial support from the Spanish Ministry MINECO and 
the European Regional development Fund (project ENE2013-48816-C5-5-R) and from 
the Regional Government of Aragon and the European Social Fund (DGA-ESF-T66 
Grupo Consolidado).
The authors are grateful to P. Launois, S. Rouziere, and C. P. Ewels for useful discussions.

\end{document}